\documentstyle[preprint,aps,epsfig,amssymb,amsbsy,amsmath,multirow]{revtex}
\newcommand{\op}[1]{%
    \fontdimen12\textfont3=2pt\fontdimen12\scriptfont3=1.4pt%
    \!\null\mathop{\vphantom{#1}\smash{#1}}\limits_{\sim}\null\!}
\newcommand{\vek}[1]{{\!\vec{\,#1}}}
\newcommand{\fmref}[1]{(\protect\ref{#1})}
\newcommand{\xref}[1]{\protect\ref{#1}}
\newcommand{\figref}[1]{Fig.~\protect\ref{#1}}

\def\ket#1{\, | \, {#1} \, \rangle}
\def\half{{\frac{1}{2}}}

\newcommand{\element}[2]{$^{#1}$#2}
\begin{document}
\tightenlines
\draft
\title{Unexpected properties of the first excited state of
non-bipartite Heisenberg spin rings}
\author{J. Schnack\cite{correspondence}}
%\footnote{jschnack\char'100uos.de,
%http://www.physik.uni-osnabrueck.de/makrosysteme/}} 
\address{Universit\"at Osnabr\"uck, Fachbereich Physik \\  
         Barbarastr. 7, 49069 Osnabr\"uck, Germany}
%\date{\today}
\maketitle
\begin{abstract}
Systematic properties of the first excited state are presented
for various ring sizes and spin quantum numbers which are only
partly covered by the theorem of Lieb, Schultz and Mattis. For
odd ring sizes the first excited energy eigenvalue shows unexpected
degeneracy and related shift quantum numbers. As a byproduct the ground
state energy as well as the energy of the first excited state of
infinite chains are calculated by extrapolating the properties
of only a few, finite, antiferromagnetically coupled Heisenberg
rings using the powerful Levin sequence acceleration method.

\end{abstract}
\widetext
%%%%%%%%%%%%%%%%%%%%%%%%%%%%%%%%%%%%%%%%%%%%%%%%%%%%%%%%%%%%%%%%%%%%%%%%
\section{Introduction and summary}

Exact diagonalization methods
\cite{BeG90,Kou97,Kou98,Wal99,BSS99} make it possible to
investigate small spin rings for various numbers $N$ of spin
sites and spin quantum numbers $s$, for instance in the
Heisenberg model. The symmetries of the isotropic Heisenberg
Hamilton operator allow to decompose the Hilbert space
${\mathcal H}$ into a set of mutually orthogonal subspaces
${\mathcal H}(S,M,k)$ according to the quantum numbers of the
total spin $S$, the total magnetic quantum number $M$ and the
translational quantum number $k$ of the cyclic shift
operator. Since these subspaces are much smaller, a complete
or partial diagonalization of the Hamilton matrix is feasible.

The interest in the Heisenberg model, which is known already for
a long time, was renewed by the successful synthesis of small
magnetic molecules, among them nearly perfect ring structures of
paramagnetic ions like Fe$^{3+}$ \cite{TDP94,CCF96,LGB97}. One can
say that in the majority of these molecules the ions couple
antiferromagnetically and the spectrum is rather well described
by the Heisenberg model with isotropic next neighbour
interaction.

Looking at the properties of the calculated spin rings one
realizes that not only the ground states, but also the first
excited states share systematic properties which are understood
only for Heisenberg spin rings of an even number of spin sites
(bipartite systems), i.e. they can in part be derived from the
theorem of Lieb, Schultz and Mattis \cite{LSM61,LiM62}.
Non-bipartite rings which have an odd number of sites, and thus
can be characterized as frustrated, show unexpected properties in
degeneracy and translational quantum number.

Knowing systematic rules for quantum numbers also of
non-bipartite systems would be very useful for comparison of
theoretical results with measurements in ESR/EPR, torque
magnetometry or neutron scattering, see
e.g. \cite{BeG90,WSK99}. One could employ knowledge about
quantum numbers of ground and first excited states in order to
understand the thermal behaviour of quantities like the magnetic
susceptibility. In addition these exact values may help to
improve low temperature approximations. Usually the high
temperature behaviour of observables is well known, e.g. from
classical spin dynamics \cite{LLB:JCP98}, but at low temperature
such approximations are poor. The knowledge of ground and first
excited states could already be sufficient for a considerable
improvement.

Having evaluated the spectra of small Heisenberg rings with
isotropic next-neighbour interaction one can approximate the
infinite chain limit, which for the $s=\half$ ground state is
known as the Bethe-Hulth\'{e}n limit \cite{Bet31,Hul38}. Because
the sequences converge rather slowly and only a limited number
of energy eigenvalues can be evaluated, the Levin $u$-sequence
acceleration method \cite{Lev73,Lub77} is employed, which leads
to impressive estimates of the antiferromagnetic ground state
energies as well as of the excitation gap for infinite rings or
chains of larger spin quantum numbers.

\section{Systematic properties of the spectrum}

The Hamilton operator of the Heisenberg model with
antiferromagnetic, isotropic
next neighbor interaction between spins of equal spin quantum
number $s$ is given by
%--------------------------------------------------------
\begin{eqnarray}
\label{E-2-1}
\op{H}
&=&
-
2\,J\,
\sum_{x=1}^N\;
\op{\vek{s}}(x) \cdot \op{\vek{s}}(x+1)
\ ,\quad \forall x: s(x)=s
\ ,\quad J < 0
\ ,\quad N+1\equiv 1
\ .
\end{eqnarray}
%--------------------------------------------------------
$\op{H}$ commutes with the total spin $\op{\vek{S}}$ and its
three-component $\op{S}^3$. In addition $\op{H}$ is invariant
under cyclic shifts generated by the cyclic shift operator
$\op{T}$. $\op{T}$ is defined by its action on the product basis
\fmref{E-2-2}
%--------------------------------------------------------
\begin{eqnarray}
\label{E-2-3}
\op{T}\,
\ket{m_1, \dots, m_{N-1}, m_N}
=
\ket{m_N, m_1, \dots, m_{N-1}}
\ ,
\end{eqnarray}
%--------------------------------------------------------
where the product basis is constructed from single-particle
eigenstates of all $\op{s}^3(x)$
%--------------------------------------------------------
\begin{eqnarray}
\label{E-2-2}
\op{s}^3(x)\,
\ket{m_1, \dots, m_x, \dots, m_N}
=
m_x\,
\ket{m_1, \dots, m_x, \dots, m_N}
\ .
\end{eqnarray}
%--------------------------------------------------------
The translational quantum number $k=0,\dots, N-1$ modulo $N$
labels the eigenvalues of $\op{T}$ which are the $N$-th roots of
unity
%--------------------------------------------------------
\begin{eqnarray}
\label{E-2-4}
z
=
\exp\left\{
-i \frac{2\pi k}{N} 
\right\}
\ .
\end{eqnarray}
%--------------------------------------------------------
Exact diagonalization methods \cite{BSS99,Wal99} allow to evaluate
eigenvalues and eigenvectors of $\op{H}$ for small spin rings of
various numbers $N$ of spin sites and spin quantum numbers
$s$. Systematic investigations \cite{BoF64,BoJ83,FLM91,Kar94,BSS99B}
revealed interesting properties of ground state quantum
numbers, compare table \xref{T-2-1}, which only for bipartite
rings can be explained by the theorem of Lieb, Schultz and Mattis
\cite{LSM61,LiM62}. 
The ground state spin quantum number and the degeneracy, for
example, depend solely on $N\!\cdot\!s$.  If $N\!\cdot\!s$ is
integer, then the ground state has $S=0$ and is non-degenerate;
if $N\!\cdot\!s$ is half integer, then the ground state has
$S=1/2$ and is fourfold degenerate \cite{BSS99B}.

It appears that for the properties of the first excited state
such rules do not hold in general, but only for ``high enough" $N$
($N>5$). Then, as can be anticipated from table \xref{T-2-1}, we
can conjecture that
\begin{itemize}
\item if $N$ is even, then the first excited state has $S=1$ and
is threefold degenerate, and
\item if $N$ is odd and the single particle spin is
half-integer, then the first excited state has $S=3/2$ and is
eightfold degenerate, whereas
\item if $N$ is odd and the single particle spin is
integer, then the first excited state has $S=1$ and is
sixfold degenerate.
\end{itemize}
Except for small odd $N$ also the cyclic shift quantum numbers
$k$ of the first excited state show an interesting regularity.
\begin{itemize}
\item For odd $N\ge 7$, $k$ assumes a certain value
for all integer spins and another value
for all half-integer spins. We conjecture that the $k$ quantum
numbers for half-integer spins are $k=3\lfloor(N+1)/4\rfloor$
and $k=N-3\lfloor(N+1)/4\rfloor$.
$\lfloor(N+1)/4\rfloor$ symbolizes the
greatest integer less or equal to $(N+1)/4$. For integer spins
numerical data are poor but it seems that $k$ is as close as
possible to $N/2$, i.e. $k=\lfloor N/2\rfloor$ and
$k=N-\lfloor N/2\rfloor$. 
\item  For even $N$ the shift quantum number
$k$ is $N/2$, if $N/2$ is also even; if $N/2$ is odd, $k=0$ for
half-integer spin and $k=N/2$ for integer spin. 
\end{itemize}
For spin-$\half$-rings these properties may be also derived
using the Bethe ansatz \cite{Bet31,Hul38}.

\section{Infinite chain limits}

Besides the importance of the above presented results for
magnetic molecules \cite{BeG90,DGP93,CCF96,PDK97,WSK99,CR4}, the
obtained energy eigenvalues enable us to estimate the
antiferromagnetic ground state energy $E_0(N)$ in the large $N$
limit for a variety of spin quantum numbers. Of course this
calculation cannot compete with nowadays DMRG results, but
reaches astonishingly close.

\begin{figure}[t]
\begin{center}
\epsfig{file=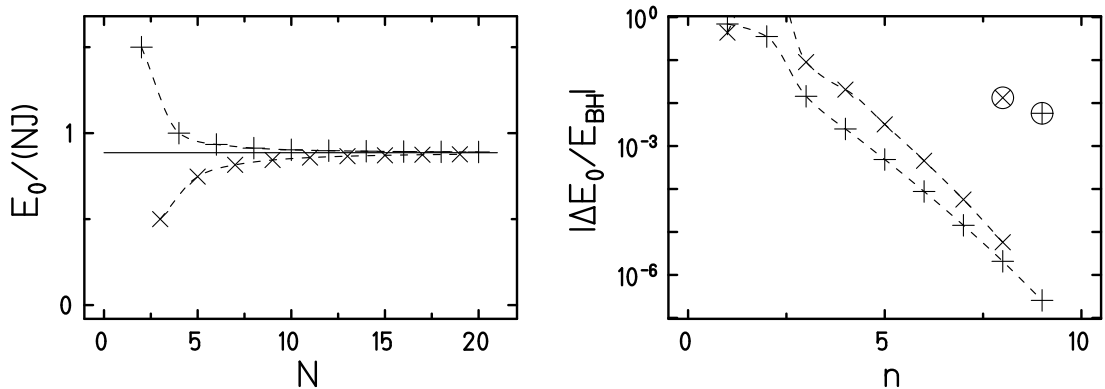,width=150mm}
\vspace*{1mm}
\caption[]{L.h.s: Ground state energies (symbols) for
antiferromagnetic coupled Heisenberg rings of $s=1/2$ compared
to the large $N$ limit of Bethe and Hulth\'{e}n (solid line).
R.h.s: Relative deviation of the Levin $u$-estimate from the
limit of Bethe and Hulth\'{e}n. Plus symbols are used for even
$N$, crosses for odd $N$. The circled symbols show how much the
ground state energies itself deviate from the limit.}  
\label{F-3-1}
\end{center} 
\end{figure}

As one can see in table \xref{T-2-1} or \figref{F-3-1} (l.h.s.)
the convergence of energy eigenvalues with $N$ is rather
slow. Therefore, an improved estimate is calculated using the
Levin $u$-sequence acceleration method
\cite{Lev73}, which is appropriate if the series elements form
an approximately linear function in $\frac{1}{i^k}$ with a
certain positive $k$, which to first approximation is the case
for the sequences of table \xref{T-2-1}. This observation is in
accordance with the Wess-Zumino-Witten model, see
e.g. \cite{AGS:JPA89}, which yields
%--------------------------------------------------------
\begin{eqnarray}
\label{E-2-8}
\frac{E_0}{N}
\approx
\epsilon_{\infty}
-
\frac{\alpha}{N^2}
\ .
\end{eqnarray}
%--------------------------------------------------------
Let us denote the elements of our series by $U_1, U_2,
U_3,\dots$. In order to construct monotonic sequences the
alternating series for a certain spin are divided into two
monotonic subseries e.g. for $s=\half$: $U_1=E_0(N=2),
U_2=E_0(N=4), U_3=E_0(N=6), \dots$. The differences between
successive sequence elements are labeled $u_1=U_1, u_2=U_2-U_1,
u_3=U_3-U_2, \dots$. Then the Levin $u$-estimate using $n$
series elements reads
%--------------------------------------------------------
\begin{eqnarray}
\label{E-2-5}
U[n]
=
\frac{\sum_{k=1}^n (-1)^{k-1}\binom{n}{k}k^{n-2}\frac{U_k}{u_k}}
     {\sum_{k=1}^n (-1)^{k-1}\binom{n}{k}k^{n-2}\frac{1}{u_k}}
\ .
\end{eqnarray}
%--------------------------------------------------------
The following example for the sequence built of the energies for
even numbers of spin $s=1/2$ demonstrates how fast the Levin
$u$-estimate approaches the correct limit
%--------------------------------------------------------
\begin{eqnarray}
\label{E-2-6}
\{
U[n]
\}
&=&
\{
1.5,1.2,0.8735510038,0.8885066176,0.8858640679,0.8863719562,
\\
&&
0.8862817068,0.8862961998,0.8862941347,\dots
\}
\rightarrow
0.8862943611
=
2 \ln 2 - \half
\nonumber
\ .
\end{eqnarray}
%--------------------------------------------------------
Roughly one can say that, if the deviations from a power-law
behaviour are small, $n$ sequence elements lead to an accuracy
of about $n-4$ digits, see \figref{F-3-1} (r.h.s.). Of course
the limit can only be as accurate as the individual sequence
elements.

\begin{figure}[t]
\begin{center}
\epsfig{file=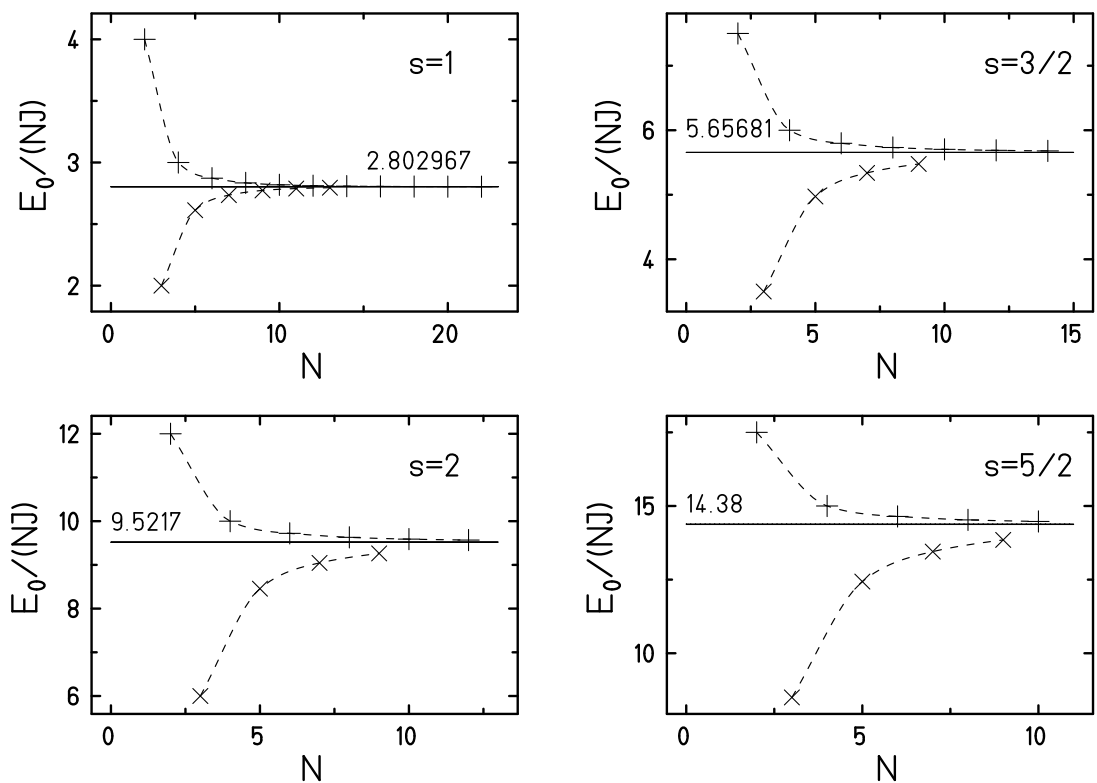,width=150mm}
\vspace*{1mm}
\caption[]{L.h.s: Ground state energies (symbols) for
antiferromagnetic coupled Heisenberg rings compared to the Levin
$u$-estimate (hatched area, sometimes shrunk to a line). Plus
symbols are used for even $N$, crosses for odd $N$.}
\label{F-3-2}
\end{center} 
\end{figure}

The procedure is also applied to spin values $s=1, 3/2, 2, 5/2$
and the results are depicted in \figref{F-3-2}. The hatched
areas indicate the interval according to the gained accuracy of
the sequence acceleration method, the attached number denotes
the mean value of the interval. Since the sequences built from
values for even $N$ converge much faster, they determine the
results. 

In the upper left of \figref{F-3-2} the estimate for the
infinite chain limit for spin $s=1$ is shown. The estimated
ground state energy of $E_0/(NJ)=2.802967\pm0.0000005$ agrees
very well with the result $2.802968077942(4)$ found in
Ref.~\cite{WhH93} and also with other DMRG and TMRG calculations, see
e.g. \cite{GJL:PRB94,Xia98}.  Also for $s=3/2$ the limit is
rather well approximated, the value of $E_0/(NJ)=5.65681\pm
0.00001$ suggests new discussion of the results
$5.65666\pm0.00002$ of Ref.~\cite{HWH:PRL96} and $5.658$ of
Ref.~\cite{Xia98}. For these sequences previously obtained exact
diagonalization results have been used, too
\cite{GlS:PRB84,PaB:PRB85,Mor:PRB87,Lin:PRB90,GJL:PRB94}. 

\begin{figure}[t]
\begin{center}
\epsfig{file=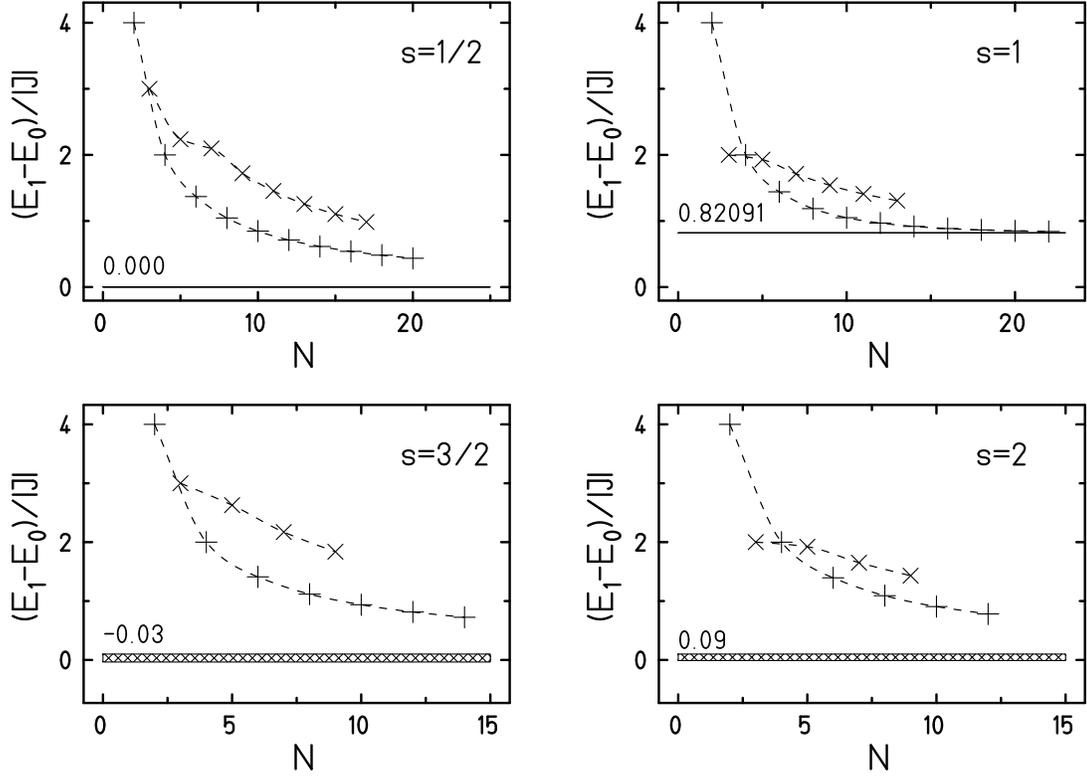,width=150mm}
\vspace*{1mm}
\caption[]{Excitation energy of the first excited state for
antiferromagnetic coupled Heisenberg rings and the Levin
$u$-estimate (hatched area). Plus symbols are used for even $N$,
crosses for odd $N$.  To generate these figures also previously
obtained exact diagonalization results of other authors have
been used 
\cite{GlS:PRB84,PaB:PRB85,Mor:PRB87,Lin:PRB90,GJL:PRB94}.  }
\label{F-3-3}
\end{center} 
\end{figure}

The excitation energies of the first excited state, see table
\xref{T-2-1}, enable us to approach the gap for infinite
chains or rings. It vanishes for half integer spins and remains
finite for integer spins, Haldane conjecture
\cite{Hal83A,Hal83B}. The following example shows the Levin $u$-sequence 
for $s=1/2$ and even $N$
%--------------------------------------------------------
\begin{eqnarray}
\label{E-2-7}
\{
U[n]
\}
&=&
\{
4.,3.,-2.464274955,0.26749212,0.05231106824,-0.04294415611,
\\
&&
0.01234278872,-0.003416797715,0.0001536344576363,\dots
\}
\rightarrow
0
\nonumber
\ .
\end{eqnarray}
%--------------------------------------------------------
In \figref{F-3-3} one can see that the behaviour of the sequence
is much smoother for even $N$, whereas the somewhat strange
behaviour for small odd $N$ destroys a fast convergence.  The
convergence of the gap sequences is slowed down by much stronger
logarithmic corrections to the power-law behaviour than present
in the ground state energy sequences. Thus the gained accuracy
for higher spin quantum numbers is rather limited and larger
rings together with methods like DMRG have to be used, see
e.g. \cite{GJL:PRB94,SGJ:PRB96,WQY:PRB99}.

%%%%%%%%%%%%%%%%%%%%%%%%%%%%%%%%%%%%%%%%%%%%%%%%%%%%%%%%%%%%%%%%%%%%%%%%
\section*{Acknowledgments}
The author would like to thank M.~Luban for introducing the
Levin methods to him and K.~B\"arwinkel, E.~Kotomin, D.~Mentrup
and H.-J.~Schmidt for helpful discussions.
%%%%%%%%%%%%%%%%%%%%%%%%%%%%%%%%%%%%%%%%%%%%%%%%%%%%%%%%%%%%%%%%%%%%%%%%

%-----------------------------------------------------------------------
\begin{table}[t]
\begin{center}
\begin{tabular}{|cc||r|r|r|r|r|r|r|r|r|l|}
\hline
&$s$&\multicolumn{9}{c|}{$N$}&\\
&& 2&3&4&5&6&7&8&9&10&\\
\hline\hline
\multirow{36}{0mm}{}
&             & 1.5 & 0.5  & 1 & 0.747 & 0.934 & 0.816 
& 0.913 & 0.844 & 0.903 & $E_0/(NJ)$\\
&$\frac{1}{2}$& 1   & 4    & 1 & 4        & 1        & 4        
& 1        & 4        & 1 & deg\\
&             & 0 &1/2& 0 & 1/2 & 0   & 1/2 & 0   & 1/2 & 0   & $S$\\
&             & 1   & 1, 2 & 0 & 1, 4     & 3        & 2, 5
& 0        & 2, 7     & 5 & $k$\\
\cline{2-12}
&             &4.0&3.0&2.0&2.236&1.369&2.098&1.045&1.722&0.846& $\Delta E/|J|$\\
&$\frac{1}{2}$& 3 & 4 & 3 &2    & 3   & 8   & 3   & 8   & 3   & deg\\
&             & 1 &3/2& 1 & 1/2 & 1   & 3/2 & 1   & 3/2 & 1   & $S$\\
&             & 0 & 0 & 2 & 0   & 0   &1, 6 & 4   & 3, 6& 0   & $k$\\
\cline{1-12}
&   & 4 & 2 & 3 & 2.612 &2.872&2.735&2.834&2.773&2.819& $E_0/(NJ)$\\
&$1$& 1 & 1 & 1 & 1     & 1   & 1   & 1   & 1   & 1   &  deg\\
&   & 0 & 0 & 0 & 0     & 0   & 0   & 0   & 0   & 0   & $S$\\
&   & 0 & 0 & 0 & 0     & 0   & 0   & 0   & 0   & 0   & $k$\\
\cline{2-12}
&   &4.0&2.0    &2.0&1.929&1.441&1.714&1.187&1.540&1.050& $\Delta E/|J|$\\
&$1$& 3 & 9     & 3 & 6   & 3   & 6   & 3   & 6   & 3   &  deg\\
&   & 1 & 1     & 1 & 1   & 1   & 1   & 1   & 1   & 1   & $S$\\
&   & 1 &0, 1, 2& 2 & 2, 3& 3   &3, 4 & 4   &4, 5 & 5   & $k$\\
\cline{1-12}
&             & 7.5 & 3.5 & 6 &4.973&5.798&5.338&5.732&5.477&$5.704^{\dagger\dagger}$& $E_0/(NJ)$\\
&$\frac{3}{2}$& 1   & 4   & 1 & 4   & 1   & 4   & 1   & 4   & 1 &  deg\\
&             & 0 &1/2& 0 & 1/2 & 0   & 1/2 & 0   & 1/2 & 0    & $S$\\
&             & 1   & 1, 2& 0 & 1, 4& 3   & 2, 5& 0   & 2, 7& 5 & $k$\\
\cline{2-12}
&             &4.0&3.0    &2.0&2.629&1.411&2.171&1.117&1.838&$0.938^{\dagger\dagger}$& $\Delta E/|J|$\\
&$\frac{3}{2}$& 3 & 16    & 3 & 8   & 3   & 8   & 3   &  8  & 3 & deg\\
&             & 1 &3/2    & 1 & 3/2 & 1   & 3/2 & 1   & 3/2 & 1 & $S$\\
&             & 0 &0, 1, 2& 2 & 2, 3& 0   & 1, 6& 4   &3, 6 & 0 & $k$\\
\cline{1-12}
&   & 12 & 6 & 10 &8.456&9.722&9.045&9.630&$9.263^{\dagger\dagger}$&$9.590^{\dagger\dagger}$& $E_0/(NJ)$\\
&$2$& 1  & 1 & 1  & 1   & 1   & 1   & 1   & 1 & 1 &  deg\\
&   & 0 & 0 & 0 & 0     & 0   & 0   & 0   & 0 & 0 & $S$\\
&   & 0  & 0 & 0  & 0   & 0   & 0   & 0   & 0 & 0 & $k$\\
\cline{2-12}
&             &4.0&2.0    &2.0&1.922&1.394&1.652&1.091&$1.431^{\dagger\dagger}$&$0.906^{\dagger\dagger}$& $\Delta E/|J|$\\
&$2$          & 3 & 9     & 3 &  6  & 3   & 6   & 3   & 6   & 3 & deg\\
&             & 1 & 1     & 1 &  1  & 1   & 1   & 1   & 1   & 1 & $S$\\
&             & 1 &0, 1, 2& 2 & 2, 3& 3   &3, 4 & 4   & 4, 5& 5 & $k$\\
\cline{1-12}
&             & 17.5 & 8.5 & 15 &12.434&14.645&13.451&$14.528^{\dagger}$&$13.848^{\dagger\dagger}$&$14.475^{\dagger\dagger}$& $E_0/(NJ)$\\
&$\frac{5}{2}$& 1    & 4   & 1  & 4    & 1    & 4    & 1 & 4 & 1 &  deg\\
&             & 0    &1/2  & 0  & 1/2  & 0    & 1/2  & 0 &1/2& 0 & $S$\\
&             & 1    & 1, 2& 0  & 1,4  & 3    & 2, 5 & 0 &2, 7& 5 & $k$\\
\hline
\end{tabular}
\vspace*{5mm}
\end{center}
\caption[]{Properties of ground and first excited state of AF
Heisenberg rings for various $N$ and $s$: ground state energy
$E_0$, gap $\Delta E$, degeneracy $deg$, total spin $S$ and
shift quantum number $k$.
$\dagger$ -- O.~Waldmann, private communication.
$\dagger\dagger$ -- projection method \cite{Man91}. 
Values for higher $N$ are available from the author.
}\label{T-2-1}
\end{table}
%----------------------------------------------------------------------- 

%===================    figure   =================================
%===================    figure   =================================

%===================    figure   =================================
%===================    figure   =================================

%===================    figure   =================================
%===================    figure   =================================

%%%%%%%%%%%%%%%%%%%%%%%%%%%%%%%%%%%%%%%%%%%%%%%%%%%%%%%%%%%%%%%%%%%%%%%%
\end{document}